\documentclass[aps,prl,twocolumn,showpacs,floatfix]{revtex4}
\usepackage{graphicx,bm,amsmath,amssymb,natbib,url,epsfig,psfrag}

\begin{document}

\title{Klein Tunneling and Berry Phase $\pi$ in Bilayer Graphene with a Band Gap}
\author{Sunghun Park}
\author{H.-S. Sim}
\affiliation{Department of Physics, Korea Advanced Institute of
Science and Technology, Daejeon 305-701, Korea}

\date{\today}

\begin{abstract}
Klein tunneling in gapless bilayer graphene, perfect reflection of electrons injecting normal to a $pn$ junction, is expected to disappear in the presence of energy band gap induced by external gates. We theoretically show that the Klein effect still exists in gapped bilayer graphene, provided that the gaps in the $n$ and $p$ regions are balanced such that the polarization of electron pseudospin has the same normal component to the bilayer plane in the regions.
We attribute the Klein effect to Berry phase $\pi$ (rather than the conventional value $2 \pi$ of bilayer graphene) and to electron-hole and time-reversal symmetries.
The Klein effect and the Berry phase $\pi$ can be identified in an electronic Veselago lens,
an important component of graphene-based electron optics.
\end{abstract}

%%We theoretically show that Klein tunneling in gapless bilayer graphene is realized in bilayer graphene
%%$n$-$p$-$n$ junction with band gap.
%%It is achieved when an angle between electron pseudospin and the bilayer plane in the $n$ region is the same as that in the $p$ region.
%Klein tunneling in gapless bilayer graphene,
%perfect reflection of electrons from a $n$-$p$ junction at zero incidence angle, may be difficult to observe in experiment,
%due to a band gap induced by external gates.
%We theoretically show that the Klein tunneling effect is realized in gapped bilayer graphene.
%%It is achieved when an angle between electron pseudospin and the bilayer plane in the $n$ region is the same as that in the $p$ region.
%It is achieved when the electrons in the $n$ and $p$ regions have the same pseudospin components normal to the bilayer plane.
%We interpret the perfect reflection by
%Berry phase $\pi$ (rather than conventional Berry phase $2 \pi$)
%and a combination of electron-hole and time-reversal symmetries.
%We discuss how to observe the Klein tunneling effect experimentally.

%Our findings enable to observe the effect in experiments.
%invariant... %potential difference


%03.65.Vf Phases: geometric; dynamic or topological
%72.80.Vp Electronic transport in graphene
%73.23.-b Electronic transport in mesoscopic systems
%73.63.-b Electronic transport in nanoscale materials and structures
%73.23.Ad Ballistic transport

%81.05.-t Specific materials: fabrication, treatment, testing, and analysis
%81.05.ue Graphene
%85.35.-p Nanoelectronic devices

%73.40.-c Electronic transport in interface structures
%73.40.Gk Tunneling
%73.40.Lq Other semiconductor-to-semiconductor contacts, p-n junctions, and heterojunctions

\maketitle

{\it Introduction.---} Klein tunneling in graphene is a striking phenomenon~\cite{Katsnelson}, analogous to the behavior of relativistic particles.
In monolayer graphene, it predicts that a low-energy electron (massless Dirac fermion) injecting normal to a potential step perfectly transmits through the step regardless of the step height.
This effect results from the chirality of the electron~\cite{Katsnelson, NetoRMP, Beenakker,  Cheianov}, i.e., the backscattering is forbidden by the orthogonality of the pseudospins of two electrons moving in the opposite direction.
It is also attributed to Berry phase $\pi$~\cite{Ando, Novoselov1, Zhang1} of the pseudospin,
as it is accompanied by a jump in reflection phase by $\pi$ around the normal injection~\cite{Shytov}.
Experimental efforts have been done to observe the perfect transmission~\cite{Gordon, Gorbachev} and the phase jump~\cite{Young}.

%
%Klein tunneling in graphene is a striking phenomenon~\cite{Katsnelson},
%analogous to the behavior of relativistic particles.
%In the low energy regime, electrons behave as massless Dirac fermions
%with Berry phase $\pi$ \cite{NetoRMP, Beenakker, Novoselov1, Zhang1},
%exhibiting perfect transmission through a $n$-$p$ junction at zero incidence angle \cite{Katsnelson, Ando, Cheianov}.
%A jump in reflection phase by $\pi$ at the $n$-$p$ junction appears with the perfect transmission,
%%The phase jump is attributed to the Berry phase,
%and manifests the Klein tunneling effect \cite{Shytov, Young}.
%To observe the perfect transmission and the phase jump,
%experimental efforts have been done in a bipolar junction \cite{Gordon, Gorbachev} and in an interferometry \cite{Young}.

On the other hand, Bernal-stacked bilayer graphene has different features from the monolayer.
Its low-energy electrons are massive Dirac fermions having pseudospin of different origin from the monolayer, showing Klein tunneling of the opposite behavior, perfect reflection of electrons injecting normal to a $pn$ junction~\cite{Katsnelson}.
However, little attention has been paid to the bilayer Klein effect, although it may play a crucial role in bilayer graphene electronics, which has attracted much attention because of the tunability of energy band gap~\cite{McCann, Ohta, Castro, Oostinga, Zhang2, Nandkishore}.
For instance, it has not been reported whether the effect survives in the presence of a band gap, which is difficult to avoid in the junction formed by external gates.
It is also meaningful to see the fundamental link of the effect with Berry phase of the bilayer~\cite{Novoselov2},
and with electron focusing in Veselago lens \cite{Veselago, CheianovLens} for electron optics.

\begin{figure}
\includegraphics[width=0.46\textwidth]{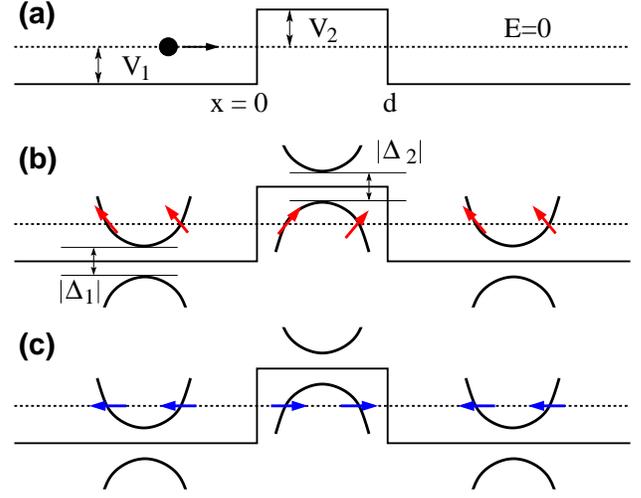}
\caption{(Color online) Electron transmission through an $npn$ junction in bilayer graphene.
The junction is formed by a potential barrier of width $d$ in (a), and has low-energy bands in (b). The potential energy and band gap in the $n$ ($p$) regions are denoted by $V_1$ and $\Delta_1$ ($V_2$ and $\Delta_2$), respectively. Red arrows represent the pseudospin polarization vector $\vec{v}$ of the electron with energy $E = 0$ (shown by dotted lines). (c) When $\Delta_1/V_1 = \Delta_2/V_2$, the rotation of electron pseudospinor by the unitary operator $U$ in Eq.~\eqref{THamiltonian} results in the orthogonal alignment of $\vec{v}$'s as blue arrows. In this case, the bilayer Klein effect (perfect reflection) occurs even with finite band gap.
}
\label{fig:setup}
\end{figure}

In this Letter, we study the Klein effect in Bernal-stacked bilayer graphene
bipolar ($npn$ or $np$) junction with band gap. % [Figs.~\ref{fig:setup} and~\ref{fig:lens}].
%, focusing on an $npn$ junction with a band gap [see Fig.~\ref{fig:setup}] and on an $np$ junction [Fig.~\ref{fig:lens}]. %; our finding also applies to a $pn$ junction.
The Klein effect is found to survive in the presence of the gap, provided that the gaps in the $n$ and $p$ regions are balanced such that electrons have the same value of $v_z$ in the regions, where $v_z$ is the component of pseudospin polarization vector [see Eq.~\eqref{def:alp}] normal to the bilayer plane.
We attribute the effect to Berry phase $\pi$ and to the electron-hole and time-reversal symmetries defined in a single valley of graphene.
The Berry phase results in the jump by $\pi$ of the transmission phase through a $pn$ (or $np$) interface around the normal incidence of electrons to the interface.
%when the angle of electron incidence to the interface varies near the normal incidence.
On the other hand, the Klein effect disappears, i.e., the transmission of electrons with normal incidence is finite, when $v_z$ differs between the $n$ and $p$ regions.
% or when the central $p$ region of the $npn$ junction is not much wider than electron wavelength.
%We discuss how to experimentally observe the effect and the Berry phase $\pi$.
We show how to detect this effect and the Berry phase $\pi$ in a Veselago lens.

{\it Setup.---} The $npn$ junction in Fig.~\ref{fig:setup} is formed by position-dependent
potential energy and a band gap,
\begin{equation} \label{Potential}
(V(x),~\Delta (x)) = \left\lbrace
      \begin{array}{ll}
        (V_1 ,~\Delta_1), \,\,\, & \hbox{$x<0$ and $x > d$},\\
        (V_2,~\Delta_2), \,\,\, & \hbox{$0<x<d$}.
      \end{array}
    \right.
\end{equation}
Assuming that $V(x)$ and $\Delta(x)$ smoothly vary in the length scale of lattice constant, we ignore the mixing between K and K$'$ valleys of the graphene.
%The smoothly varying potential barrier will be considered later.
An electron state $\Psi$ with low energy $E$ in the K valley is governed by the massive Dirac Hamiltonian~\cite{McCann} of the junction, % of $H \Psi = E \Psi$,
\begin{equation}
H = \frac{v^2}{\gamma} \vec{\sigma} \cdot \left( -p^{2}_{x} + p^{2}_{y}, - 2 p_{x} p_{y}, \frac{\gamma \Delta(x)}{2 v^2} \right) + V (x), \label{Hamiltonian}
\end{equation}
and the K$'$ valley is described by a similar Hamiltonian.
$\Psi$ has the pseudospin components describing the lattice sites A$_1$ and B$_2$ of the bilayer,
where A$_l$ and B$_l$ denote the two basis sites of layer $l=1,2$. $\vec{\sigma} = (\sigma_x, \sigma_y, \sigma_z)$ is the pseudospin Pauli operator, $\vec{p} = (p_x, p_y)$ is the momentum measured relative to the valley center, $\gamma \approx 390 \, \, \textrm{meV}$ is the interlayer (B$_1$-A$_2$) coupling, and $v \approx 10^6$~m/s.
We assume the low-energy regime of $|E-V_j|,~|\Delta_j| \ll \gamma$,
ignore the trigonal warping~\cite{McCann} by assuming $|E-V_j|,~|\Delta_j| > 0.005 \gamma$,
and also ignore the case of $|V_{j=1,2}| < |\Delta_{j}/2|$ [i.e., $|\alpha_j| > 2$ in Eq.~\eqref{def:alp}] in which electron states with $E$ are evanescent waves inside the band gaps. Without loss of generality, we set $E=0$ hereafter; see Fig.~\ref{fig:setup}(b).

We calculate the transmission coefficient $t$ and probability $T = |t|^2$ of an electron with $E=0$ through the junction, as a function of its incidence angle $\theta_1$ to the interface at $x = 0$; $\theta_1 = 0$ at the normal incidence. We below consider electrons in the K valley only, as the K$'$ valley shows the same result; while $t$ is valley dependent at finite $\theta_1$ in a $pn$ junction~\cite{Schomerus}, it is independent in the $npn$ junction in Fig.~\ref{fig:setup} because of its left-right symmetry with respect to $x=d/2$.
The electron has the wave function of $\Psi(x,y)=e^{i k_y y} \psi(x)$ due to the translation invariance along $\hat{y}$, where $k_y$ is the wave vector along $\hat{y}$ and $\psi(x)$ is written as a superposition of propagating and evanescent waves.
The continuity of $\psi(x)$ and $d\psi(x)/dx$ at $x=0,d$ under Eq.~\eqref{Hamiltonian} determines $t$~\cite{Katsnelson}.

%To understand the Klein effect,
For further discussion, it is useful to see the pseudospin polarization vector $\vec{v}_j \equiv \langle \Psi_j | \vec{\sigma} | \Psi_j \rangle$ of the electron, where $\Psi_j$ is the wave function $\Psi$ in the region $j$ of $\Delta_j$. While $\vec{v}_j$ is parallel to the bilayer ($xy$) plane (i.e., $v_z = 0$) in the case of zero band gap, one can show from Eq.~\eqref{Hamiltonian} that for finite gap it has a finite value of $v_z$ as
\begin{gather}
\vec{v}_j = -s_j \sqrt{1 - \left( \frac{\alpha_j}{2} \right)^2} \left( \textrm{cos} 2\theta_j \ \hat{x} + \textrm{sin} 2\theta_j \ \hat{y} \right) - \frac{\alpha_j}{2}\ \hat{z}, \nonumber \\
\alpha_j \equiv \Delta_{j}/V_{j}, \,\,\,\,\,\, s_j \equiv -\textrm{sgn}(V_j),
\label{def:alp}
\end{gather}
where propagation angle $\theta_j \equiv \tan^{-1} p_{y,j}/p_{x,j}$ in region $j$.

\begin{figure}
\includegraphics[width=0.45\textwidth]{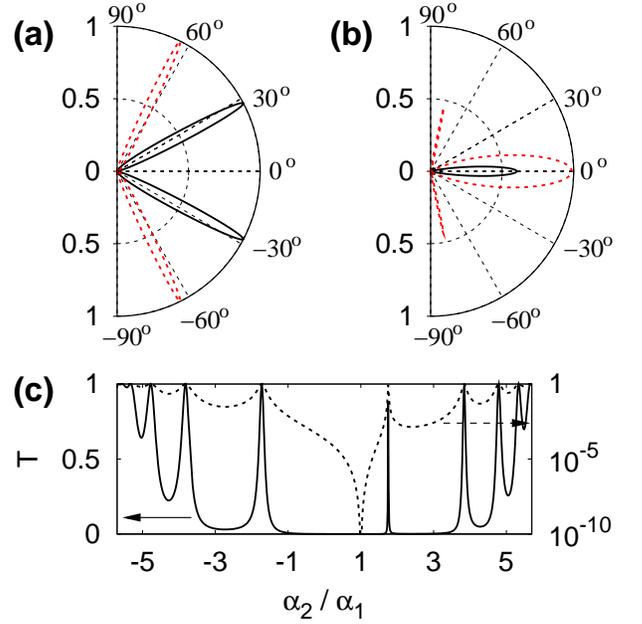}
\caption{(Color online) (a,b) Transmission probability $T = |t|^2$ of an electron through the junction as a function of electron incidence angle $\theta_1$ for the cases of (a) $\alpha_1 = \alpha_2$ and (b) $\alpha_1 \ne \alpha_2$. We choose $(\alpha_1, \alpha_2)$ as ($-0.3, -0.3$) [black solid curve] and ($-1, -1$) [red dashed] in (a),
and ($-0.3, 0.6$) [black solid] and ($-0.3, -1.6$) [red dashed] in (b).
(c) $T$ is drawn for normal incidence of $\theta_1 = 0$, with $\alpha_1 = -0.3$ and varying $\alpha_2$;
%as a function of $\alpha_2/ \alpha_1$ and with $\alpha_1 = -0.3$;
see also its log-scale plot (dashed line) on the right side.
For (a)-(c), we choose $V_1 = -0.05 \gamma$, $V_2 = 0.26 \gamma$, %the barrier width
$d = 60 l_0$ ($\gg \lambda$), and $l_0 \equiv \hbar v / \gamma \approx 1.7 \textrm{nm}$.
}
\label{fig:Trans}
\end{figure}

As shown in Fig.~\ref{fig:Trans}(a), we find that the Klein effect survives ($t = 0$ at $\theta_1 = 0$, irrespective of $V_j$ and $\Delta_j$) even in the case of finite gap, provided $\alpha_1 = \alpha_2$ (i.e., $v_{z,j=1} = v_{z,2}$) and $d \gg$ electron wavelength $\lambda$; the condition of $d \gg \lambda$ suppresses electron tunneling through the central $p$ region with width $d$, and $\lambda =  2 \pi \hbar v / \sqrt{\gamma \sqrt{V^{2}_{2} - \Delta^{2}_{2}/4}}$ in the $p$ region.
Notice that the condition of $\alpha_1 = \alpha_2$ includes the zero-gap case of $\Delta_{j=1,2} = 0$. The Klein effect also occurs in a $pn$ ($np$) interface, provided $\alpha_1 = \alpha_2$.
%that the $n$ and $p$ regions have the same value of $\alpha$.

%
%$v = \sqrt{3} a \gamma_0 / (2 \hbar) \approx 10^6 m/s$,
%$\gamma_0$ is the intralayer coupling strength between the two sites $A_i$ and $B_i$.
%

%The transmission prbability $T = |t|^2$ through the $n$-$p$-$n$ (or $p$-$n$-$p$) junctions
%is strongly depends on the values of $\delta \alpha (\equiv \alpha_2 - \alpha_1$) and $d$.
% which will be discussed below.

%
%Fig.2: Note that the wavelength $\lambda$ of the electron in the $p$ region varies from $12 l_0$ to $16 l_0$ when $|\alpha_2|$ changes from $0.3$ to $1.6$.

%We first examine the electron transmission $T$ through the $n$-$p$-$n$ junctions
%for the case of $\delta \alpha (\equiv \alpha_2 - \alpha_1 ) = 0$, $d \gg 2 \pi/ k_2$ (where
%the tunneling through the evanescent waves is negligible), and then that of finite $\delta \alpha$ or $d \lesssim 2 \pi/ k_2$.
%Here, $k_2 = \sqrt{(\gamma/(\hbar^2 v^2)) \sqrt{V^{2}_{2} - \Delta^{2}_{2}/4}}$ is the absolute value of the wave vector
%inside the barrier.

The Klein effect in the zero-gap case of $\Delta_{j=1,2} = 0$~\cite{Katsnelson} can be understood from the fact that for electrons with normal incidence of $\theta_1 =0$, the pseudospin polarization vectors $\vec{v}$ of the $n$ and $p$ regions are orthogonal to each other, resulting in the perfect reflection of $t=0$. In the case of finite gap with $\alpha_1 = \alpha_2$, however, naive application of this argument cannot explain the Klein effect, as $\vec{v}_1$ is not orthogonal to $\vec{v}_2$ as shown in Fig.~\ref{fig:setup}(b). We find that when $\alpha_1 = \alpha_2$, there exists a $2 \times 2$ unitary operator $U$ rotating electron pseudospinor such that the resulting polarization vectors $\vec{v}$'s in the $n$ and $p$ regions are orthogonal to each other; see Fig.~\ref{fig:setup}(c).
Mathematically, after the pseudospinor rotation by $U$, the continuity equations of $\psi$ and $d\psi/dx$ at $x=0,d$ become identical to those of the zero-gap case of $\Delta_{j=1,2} = 0$, which guarantees the Klein effect in the $\alpha_1 = \alpha_2$ case. Physically, the rotation by $U$ allows us to explain the Klein effect by Berry phase $\pi$ at the interface of $x=0$ (or at $d$), as below.

To see the connection to the Berry phase, we rotate the Hamiltonian $H$ by $U$ at $E=0$ in the $\alpha_1 = \alpha_2$ case,
\begin{gather}\label{THamiltonian}
\mathcal{H} = U H U^{\dagger} =
\frac{s v^2}{\gamma \sqrt{1-\alpha_{1}^{2}/4}} \vec{\sigma} \cdot \vec{q} + V (x), \\
\vec{q}= s (-p^{2}_{x} + p^{2}_{y}, -2 p_{x} p_{y},0). \nonumber
\end{gather}
We here write the expression of $U$, $U = \hat{I} \text{cos}(\phi/2) - i~\vec{\sigma} \cdot \vec{n} \, \text{sin}(\phi/2)$, where $\phi= - \text{arcsin}(\alpha_1/2)$, $\vec{n}= s ( 2 p_{x} p_{y}, -p^{2}_{x} + p^{2}_{y},0)/|p^{2}_{x} + p^{2}_{y}|$, $\hat{I}$ is the identity, and $s = +1 (-1)$ for the electron (hole)-like band. The term $\vec{\sigma} \cdot \vec{q}$ in $\mathcal{H}$ causes the chiral property that the pseudospin of $U \Psi$ is parallel to $\vec{q}$ in both electron-like and hole-like bands; this property of the same chirality between electron and hole bands is introduced by the sign factor $s$ in our definition of $\vec{q}$. Because of the chirality, the pseudospin rotates following the rotation of $\vec{q}$ during the scattering process at the interfaces of $x=0,d$, and the vectors $\vec{q}$ constitute the parameter space for the Berry phase associated to the pseudospin of the electron propagating with real $\vec{p}$.
Note that $\mathcal{H}$ has the same form as the Hamiltonian $H$ of the zero-gap case, and that $\mathcal{H}$ should not be interpreted as a Hamiltonian, since the rotation is performed at the given energy $E=0$.

\begin{figure}
\includegraphics[width=0.47\textwidth]{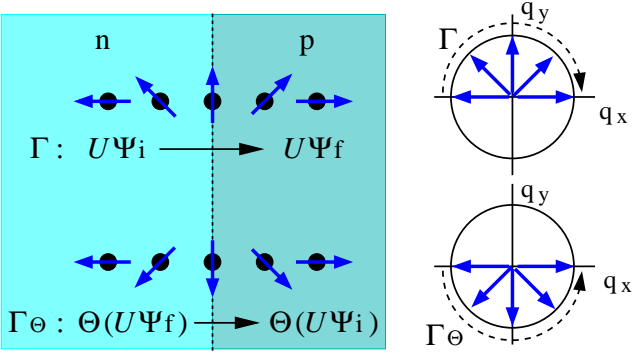}
\caption{
(Color online) Evolution of $\vec{q}$ (left panel) and pseudospin (right) of electrons injecting normal to an $np$ junction with $\alpha_1 = \alpha_2$. There are two such degenerate states $U \Psi_i$ and $\Theta U \Psi_f$ because of the symmetry $\Theta$ of $\mathcal{H}$. The pseudospin and $\vec{q}$ of the states are drawn by blue thick arrows, and they are parallel to each other due to the chirality caused by the term $\vec{\sigma} \cdot \vec{q}$ in $\mathcal{H}$. The evolution from $U \Psi_i$ to its transmitted state $U\Psi_f$ is denoted by $\Gamma$, and its symmetric process from $\Theta U \Psi_f$ to $\Theta U \Psi_i$ by $\Gamma_\Theta$. The pseudospin evolution along $\Gamma$ differs from that along $\Gamma_\Theta$ by $2 \pi$ rotation, giving rise to Berry phase $\pi$. The resulting destructive interference between $\Gamma$ and $\Gamma_\Theta$ causes the perfect reflection, the bilayer Klein effect.
}
\label{fig:Qspace}
\end{figure}

Another important feature of the $\alpha_1=\alpha_2$ case is the symmetry of $\mathcal{H}$ described by the antiunitary operator $\Theta$,
\begin{equation}
\Theta = i \sigma_z \sigma_y \mathcal{C}.
\label{CToperator}
\end{equation}
Here, $\sigma_z$ is the electron-hole conversion operator~\cite{Beenakker}, $i \sigma_y \mathcal{C}$ is the effective time-reversal operator~\cite{Beenakker}, transforming $(\vec{p},\vec{\sigma})$ to $(-\vec{p},-\vec{\sigma})$ and defined in a single valley, and $\mathcal{C}$ is the complex conjugation operator. It is a good symmetry, $[\Theta, \mathcal{H}] = 0$, and plays an important role as below.
Note that $\Theta$ has not been discussed in literature.
%where electron-hole conversion arises.
%for propagating states with real $\vec{p}$.
%(for $0.005 \gamma < |V_{j}|,~|\Delta_{j}| \ll \gamma$),

Now, combining the chirality and the symmetry in $\mathcal{H}$, we are able to connect the Klein effect with Berry phase $\pi$; see Fig.~\ref{fig:Qspace}. We consider an incident (electron-like) state $U\Psi_i$ of $\mathcal{H}$ which injects normal to the $np$ interface with $\alpha_1=\alpha_2$ at $E=0$, and its transmitted (hole-like) state $U\Psi_f$. The evolution from $U \Psi_i$ to $U \Psi_f$ is denoted by $\Gamma$. The vector $q$ of the state rotates by angle $\pi$ (clockwise or counterclockwise) due to the electron-hole conversion during the evolution, and the pseudospin of the state rotates following the rotation of $\vec{q}$ due to the chirality. On the other hand, the symmetry $\Theta$ guarantees that there is another degenerate state injecting normal to the interface, and that it evolves from $\Theta U \Psi_f$ to $\Theta U \Psi_i$ along the reversal process denoted by $\Gamma_\Theta$, in which the rotation of $\vec{q}$ and pseudospin is opposite to that of $\Gamma$. The evolution of pseudospin along $\Gamma$ differs from that along $\Gamma_\Theta$ by $2 \pi$ rotation, i.e., the difference $\Gamma - \Gamma_\Theta$ forms a loop encircling once the origin of the $\vec{q}$ space, which causes Berry phase $\pi$.
The Berry phase $\pi$ makes the destructive interference between $\Gamma$ and $\Gamma_\Theta$, hence resulting in the perfect reflection of $t_{np}=0$ at the interface, the bilayer Klein effect. Here, $t_{np}$ is the transmission coefficient through the interface.

In addition to the perfect reflection, the Berry phase also induces the phase jump of $t_{np}(\theta_1)$ by $\pi$ around normal incidence of $\theta_1=0$.
We derive $t_{np}$ around $\theta_1 \simeq 0$,
\begin{equation}
t_{np}(\theta_1) \simeq t_{0} \sin \theta_1. \label{Tstep}
\end{equation}
Here, $t_{0} = \frac{2 e_1^{1/4} (-\sqrt{e_1} + i \sqrt{e_2})}{e_2^{1/4} (\sqrt{e_1} + i \sqrt{e_2})}$ and $e_j = \sqrt{V_j^2 - \frac{\Delta_{j}^{2}}{4}}$.
%$k_j = \sqrt{(\gamma/(\hbar^2 v^2)) |V_j| \sqrt{1 - \alpha_{j}^{2}/4}}$
This result is understood by considering a state injecting to the interface with incidence angle of $\theta_1 = - \delta$, $\delta \to 0^+$. The evolution of this state approximately follows $\Gamma$. There is another state with $\theta_1 = \delta$ following $\Gamma_\Theta$. As $\theta_1$ varies from $-\delta$ to $\delta$, the evolution changes from $\Gamma$ to $\Gamma_{\Theta}$, hence the phase of $t_{np}$ jumps by the Berry phase $\pi$ at $\theta_1 = 0$.

Next, we discuss the $\alpha_1 \neq \alpha_2$ case.
In this case, the symmetry argument in Fig.~\ref{fig:Qspace} is not applicable,
and $T = |t|^2$ is nonzero at $\theta_1 = 0$; see Fig.~\ref{fig:Trans}(b) and (c).
To see more details, we derive $t(\theta_1)$ at $\theta_1 =0$ in an $npn$ junction with
$e^{-2 \pi d/\lambda},|\alpha_2 - \alpha_1| \ll 1$,
%As $|\delta \alpha|$ or $1/d$ increases, the transmission probability for $\theta_1 =0$ increases, as
%shown in the following analytic solution for transmission coefficient $t(\theta_1 =0)$,
\begin{equation}
t(0) = t_K (\delta \alpha)^2 + t_d e^{-\frac{2 \pi d}{\lambda}} + \mathcal{O}(\delta \alpha^3) + \mathcal{O}(\delta \alpha e^{-\frac{2 \pi d}{\lambda}} ).
\end{equation}
Here $\delta \alpha \equiv \alpha_2 - \alpha_1$, $t_K  =  \frac{4 \sqrt{e_1 e_2} (\sqrt{e_{1}} + \sqrt{e_{2}})^2  e^{2\pi i d/ \lambda} / (4 - \alpha_{1}^2)^2} {e^{4\pi i d/\lambda} (\sqrt{e_{1}} +i \sqrt{e_{2}})^4 - (e_{1} + e_{2})^2}$, and
$t_d = \frac{4 i \sqrt{e_{1} e_{2}}}{(\sqrt{e_{1}} + i \sqrt{e_{2}})^2}$.
%\begin{equation}
%t_0 \simeq \frac{4 i k_{1 x} k_{2 x}}{(k_{1 x} -i k_{2 x})^2} e^{-|k_{2 x}| d}
%+ \ \mathcal{O}(e^{-|k_{2 x}| d} \cdot \delta \alpha), \nonumber
%\end{equation}
%\begin{eqnarray}
%t_1 & \simeq & \frac{4 k_{1 x} k_{2 x}(k_{1 x} - k_{2 x})^2 / (4 - \alpha_{1}^2)^2}
%{e^{-i k_{2 x}d} (k_{1 x} -i k_{2 x})^4 - e^{i k_{2 x}d} (k_{1 x}^2 + k_{2 x}^2)^2} (\delta \alpha)^2 \nonumber\\
%&& + \ \mathcal{O}((\delta \alpha)^3). \nonumber
%\end{eqnarray}
The first term of $t(0)$ shows the suppression of the Klein effect in the case of $\alpha_1 \ne \alpha_2$. It oscillates with $d/\lambda$. The second term comes from the tunneling via the evanescent waves in the central $p$ region.
%when $d$ is not much smaller than $\lambda$.

We compare the bipolar junction with a bilayer monopolar ($nn'$ or $pp'$) junction.
The monopolar case has different symmetry from $\Theta$~\cite{Park}.
So, electrons injecting normal to the monopolar junction is not perfectly reflected, regardless of $\alpha_1$ and $\alpha_2$.
%We note that an electron tunneling through a monopolar ($nn'$ or $pp'$) junction~\cite{Park} with $\alpha_1 = \alpha_2$
%does not show the feature of the Klein effect.
%The electron injecting normal to the monopolar junction is not perfectly reflected,
%and there is no phase jump in its transmission amplitude around $\theta_1 \simeq 0$.
%The reason is that the symmetry defined in \eqref{CToperator} is not valid in the tunneling process, since the electron-hole conversion does not arise at the junction.

{\it Electronic Veselago lens.---} We examine spatial interference pattern
%diverging from a point source
produced by negative refraction of waves in the $np$ junction.
%We below demonstrate that an interference pattern produced by the negetive refraction
%of the waves at the junction is quite different from that in graphene \cite{CheianovLens} due to the effects.
The pattern will provide the direct evidence of the Klein effect and the Berry phase $\pi$,
and can be probed by scanning tunneling microscopy.

\begin{figure}
\includegraphics[width=0.47\textwidth]{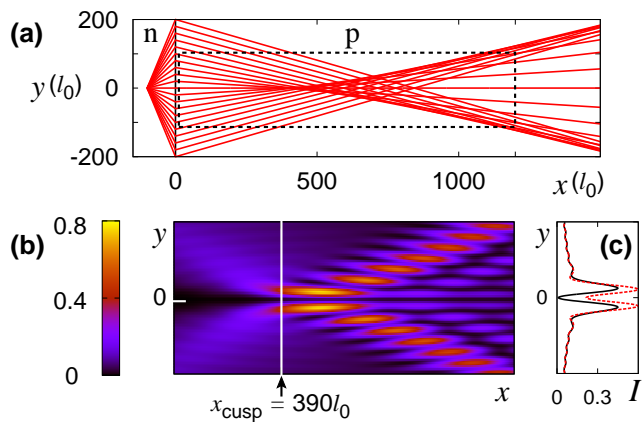}
\caption{(Color online) Veselago lens.
(a) Schematics of electron propagation in a clean $np$ junction, when an electron point source is located at
($x$,~$y$) = ($-100l_0$,~$0$) in the $n$ region.
Negative refraction occurs at the junction ($x = 0$).
%The electrons are negatively refracted at the junction ($x = 0$).
(b) Wave intensity $I(x,y)$ in the region marked by the dashed box in (a).
%Due to the Klein effect and the Berry phase $\pi$, $I(x,0) \approx 0$ along the line of $y=0$.
(c) $I(x_{{\rm cusp}},y)$ along the white line at $x = x_{{\rm cusp}}$ in (b).
%, is drawn by black solid line as a function of $y$.
%as a function of $y$, where $x$ = $x_{{\rm cusp}}$ correponds to a white line in (b).
In (a)-(c), we choose  $\alpha_1=-0.3, V_1 = -0.02 \gamma, V_2 = 0.3 \gamma$.
$\alpha_2/\alpha_1 = 1$ for (a), (b), and the black solid line in (c),
while $\alpha_2 / \alpha_1= 3$ for the red dashed line in (c), for comparison.
%In (c), $I(x_{{\rm cusp}},y)$ is also shown as the red dashed line for $\alpha_2 / \alpha_1= 3$.
%In this case, the Klein effect disappears, and $I(x_{{\rm cusp}},0)$ is finite.
}
\label{fig:lens}
\end{figure}

%We consider an electron point source \cite{CheianovLens} at ($x$,~$y$) = ($-100 l_0$,~$0$) in the $n$ region.
%The classical trajectories of the electrons propagating from $n$ to $p$ region are drawn in Fig.~\ref{fig:lens}(a).
For an electronic Veselago lens in a clean $np$ junction~\cite{CheianovLens} [see Fig.~\ref{fig:lens}], we calculate the intensity $I(x,y)= I_{{\rm K}} +I_{{\rm K}'} $ of the refracted waves in the $p$ region.
$I_{{\rm K}({\rm K}')} = |\Psi^{{\rm K}({\rm K}')}_{{\rm rfr}}|^2$, and
$\Psi^{{\rm K}({\rm K}')}_{{\rm rfr}}$ is the refracted waves of K (K$'$) valley;
for the details, see Ref.~\cite{Supplement}.

%We calculate an intensity of refracted waves in the $p$ region generated by
%an electron source located at ($x$,~$y$) = ($-100 l_0$,~$0$) in the $n$ region;
%classical trajectories of the propagating electrons are drawn in Fig.~\ref{fig:lens}(a).
%The wave function diverging from the source $\Psi^{\pm}_{\rm{source}}$ is described by out-going wave of angular momentum zero.
%
%of the Hamiltonian in Eq.~\eqref{Hamiltonian}
%in a cylindrical coordinate ($\rho,\phi$),
%\begin{equation}\label{SourceWave}
%\Psi^{\pm}_{\rm{source}}(\rho,\phi)=
%\left( \begin{array}{c}
%(\hbar^2 v^2 / \gamma) k^2~H^{(1)}_{1}(k\rho) e^{\mp i \phi} \\
%(V_1 + \Delta_1 /2) H^{(1)}_{1}(k\rho) e^{\pm i \phi}
%\end{array}
%\right),
%\end{equation}
%
%$\left( \Psi^{\pm}_{\rm{source}}(\rho,\phi) \right)^{\dagger}=\left(
%(\hbar^2 v^2 / \gamma) k^2~H^{(2)}_{1}(k\rho) e^{\pm i \phi},
%(V_1 + \Delta_1 /2) H^{(2)}_{1}(k\rho) e^{\mp i \phi}
%\right)$,
%where the sign $+$ ($-$) corresponds to the K (K$'$) valley, $\hbar v k = \sqrt{\gamma \sqrt{V^{2}_{1} - \Delta^{2}_{1}/4}}$, $H^{(2)}_{1}(k\rho)$ is the Hankel function of the second kind, and
%$(\rho~{\rm cos}~\phi, \rho~{\rm sin}~\phi) = (x + 100 l_0, y)$.
%
%The plane-wave expansion of $\Psi^{\pm}_{\rm{source}}$ and the continuity of the plane waves
%and their derivatives at $x = 0$ determine the refracted waves and the intensities $I_{{\rm K}({\rm K}')}(x,y)
%= |\Psi_{{\rm K}({\rm K}')}(x,y)|^2$. Here, $\Psi_{{\rm K}({\rm K}')}(x,y)$ is
%the refracted waves in the $p$ region;
%see supplemental material for more detail.

% for each valley.

In Fig.~\ref{fig:lens}(b), we plot $I(x,y)$ near a cusp point $(x, y) = (x_{{\rm cusp}}, 0)$,
where the caustic curve of the classical trajectories is singular;
see Fig.~\ref{fig:lens}(a).
%the trajectories are drawn in Fig.~\ref{fig:lens}(a).
%with the values of $V_1 = -0.02 \gamma$, $V_2 = 0.3 \gamma$, and $\alpha_1 = \alpha_2 = -0.3$.
%Here, the calculations are done for $\alpha_1 / \alpha_2 = 1$ and $|V_1/V_2| \approx 0.07 \ll 1$.
% and $|V_1 / V_2 | \simeq 0.07$.
%where the pseudospin directions of
%the electrons in $p$ region are almost parallel to each other. In this regime,
%the pseudospin degree of freedom is irrelevant to the interference produced by the negative refraction of the electrons.
%So, the factors of the refracted wave, $t_{np}$ in Eq.~\eqref{Tstep} and dynamical phase acquired by the electron propagation,
%only contributes to the interference.
We find that $I(x,0)$ along the line of $y = 0$ is zero, even at $x = x_{{\rm cusp}}$.
%This result is a direct consequence of the Klein effect and the Berry phase $\pi$.
For $0 < x < x_{{\rm cusp}}$, the Klein effect of $t_{np}(0) = 0$ gives $I(x,0) \propto | t_{np}(0) |^2 = 0$; see Eq.~\eqref{Tstep}.
For $x \ge x_{{\rm cusp}}$, the intensity is determined by the interference of the three refracted waves~\cite{Supplement} propagating from the source to $(x,0)$ with incidence angles $\theta_1 = 0, \theta_x,$ and $-\theta_x$, respectively, at the junction interface.
%of incidence angles $\theta_1 = 0, \theta_x,$ and $-\theta_x$ propagating from the source to ($x, 0$).
In the limit of $|V_1 / V_2| \ll 1$, the intensity is approximately given by
$I(x,0) \propto | e^{i \varphi_{dy}(0)} t_{np}(0)
+ e^{i \varphi_{dy}(\theta_x)} ( t_{np}(\theta_x) + t_{np}(- \theta_x) ) |^2 +
\mathcal{O}(|t_{np}(\theta_x)|^2 |V_1/V_2|)$, where
$\varphi_{dy}(\theta_1)$ is the dynamical phase of the refracted wave with incidence angle $\theta_1$ to the interface.
% and the factor $4$ comes from the valley and spin degeneracy.
%of the electrons propagating from the source to the point ($x,0$).
In this case, $I(x,0) \approx 0$, because of $t_{np}(0) = 0$ and 
the destructive interference coming from
the Berry phase $\pi$ effect of $t_{np}(\theta_x) = - t_{np}(- \theta_x)$;
for comparison, Fig.~\ref{fig:lens}(c) shows the $\alpha_1 \ne \alpha_2$ case where the Klein effect disappears.
The result of $I(x,y=0) = 0$ is a clear manifestation of the Klein effect and the Berry phase $\pi$.
%and is robust against a smooth $np$ junction due to $\Theta$.
This result is contrary to the maximum intensity at the cusp point in the cases of a monolayer graphene $np$ junction~\cite{CheianovLens} and of geometrical optics with negative refractive index.
%where the maximum intensity at $x = x_{{\rm cusp}}$ is achieved.
We also discuss another way of detecting the Klein effect in
an armchair nanoribbon in Ref.~\cite{Supplement}.
%Here, $x_{{\rm cusp}}= \sqrt{|V_2/V_1|}100 l_0 \approx 390 l_0$.

%Although $T$ oscillates as $d$ and $k_2$ are varied,
%$t_0$ is zero when $\delta \alpha = 0$.

%Realization of the potential energy and band gap configuration discussed above
%may be easier than the zero band gap setup in experiment.

{\it Conclusion.---} We have studied the Klein effect in bilayer graphene.
The effect can survive in the presence of energy band gap, and can be identified in
a Veselago lens.
We summarize our findings, by comparing the Klein effect in bilayer graphene and that in monolayer graphene. Both the effects result from Berry phase $\pi$ associated with pseudospin (but of different origin). The correspondence between them is found in the chirality ($\vec{\sigma} \cdot \vec{p}$ for monolayer; $\vec{\sigma} \cdot \vec{q}$ for bilayer), the parameter space of the Berry phase ($\vec{p}$; $\vec{q}$), symmetry (time reversal; $\Theta$ in Eq.~\eqref{CToperator}), and destructive interference and phase jump $\pi$ (in reflection amplitude; in transmission) at normal incidence to the junction.
Our findings will play an important role in graphene-based electron optics.

Note that, in experiments, it would be easier to study the Klein effect with a finite band gap than that of the zero-gap limit, since an $npn$ (or $np$) junction can be realized by three (two) parallel gate electrodes, while the latter requires three (two) pairs of top and bottom gates.

This work is supported by NRF (2009-0078437).

\newpage

\appendix

\section{Supplemental material for `` Klein Tunneling and Berry Phase $\pi$ in Bilayer Graphene with a Band Gap '' }

In this material, we discuss the details (the calculation method and the properties) of the Veselago lens. We also discuss how to detect the Klein effect in a bilayer graphene nanoribbon.

%\section{I. ~ Wave intensity in the Veselago lens}
\section{I.~ Electronic Veselago lens in bilayer graphene}

In this section, we provide the method of calculating the properties of the Veselago lens, and also more information about the Veselago lens.

The wave intensity in the Veselago setup (a bilayer graphene $np$ junction) is obtained,
using the plane-wave expansion method \cite{Cincotti}.
The setup is formed by the potential energy $V(x)$ and band gap $\Delta (x)$,
\begin{equation}
(V(x),~\Delta (x)) = \left\lbrace
      \begin{array}{ll}
        (V_1 ,~\Delta_1), \,\,\, & \hbox{$x<0$},\\
        (V_2,~\Delta_2), \,\,\, & \hbox{$x>0$}.
      \end{array}
    \right. \nonumber
\end{equation}
%using a plane-wave expansion method \cite{Cincotti}.
%We denote the waves diverging from an electron point source and the refracted waves by
%$\Psi^{{\rm K}({\rm K}')}_{{\rm src}}$ and $\Psi^{{\rm K}({\rm K}')}_{{\rm rfr}}$, respectively.
The electron point source is located at $(x, y) = (x_{{\rm src}},0) = (-100 l_0, 0)$ in the $n$ region, where
$l_0 = \hbar v / \gamma \approx 1.7 \textrm{nm}$.
The wave diverging from the source, $\Psi^{{\rm K}({\rm K}')}_{{\rm src}}$,
is described by the out-going wave of angular momentum zero,
\begin{equation}\label{SrcWave}
\Psi^{{\rm K}({\rm K}')}_{\rm{src}}(x,y)
=
%\frac{H^{(1)}_{1}\left( \rho \sqrt{k^{2}_{1_x} + k^{2}_{y}} \right)}{\sqrt{N_1}}
%\left( \begin{array}{c}
% \sqrt{V^{2}_1 - (\Delta^{2}_1 /4)}~ e^{\mp i \phi} \\
%(V_1 + \Delta_1 /2)~ e^{\pm i \phi}
%\end{array}
%\right),
\frac{H^{(1)}_{1} ( \rho k_1 )}{\sqrt{N_1}}
\left( \begin{array}{c}
 \sqrt{V^{2}_1 - (\Delta^{2}_1 /4)}~ e^{\mp i \phi} \\
(V_1 + \Delta_1 /2)~ e^{\pm i \phi}
\end{array}
\right). \nonumber
\end{equation}
Here, the upper (lower) sign corresponds to K (K$'$) valley,
we used the cylindrical coordinate $(\rho, \phi)$ whose origin is at $(x_{{\rm src}},0)$,
$(\rho~{\rm cos}~\phi, \rho~{\rm sin}~\phi) = (x - x_{{\rm src}}, y)$,
$H^{(1)}_{1}$ is the Hankel function of the first kind, $k_1 = \sqrt{(\gamma/ \hbar^2 v^2) \sqrt{V^{2}_{1} - \Delta^{2}_{1}/4}}$ is the wave vector of the wave, and $N_1 = 2 V^{2}_1 + V_1 \Delta_1$ is the normalization constant of the pseudospin part.
The energy $E$ of the wave, related to $k_1$ by $(E-V_1)^2 = \Delta_1^2 / 4 + (\hbar^2 v^2 k_1^2/\gamma)^2$, is set to be $E=0$ as in the main text.

In the calculation of wave refraction, we apply the following approximation.
In the spatial region where $\rho k_1 \gg 1$ is satisfied, $\Psi^{{\rm K}({\rm K}')}_{{\rm src}}$
can be approximately written \cite{Cincotti} in terms of plane waves,
\begin{widetext}
\begin{equation}
\Psi^{{\rm K}({\rm K}')}_{{\rm src}}(x,y)
\approx
\frac{-i}{\pi \sqrt{N_1}} \int^{\pi/2}_{-\pi/2} d \theta_1~ e^{i k_{1_x}(x - x_{{\rm src}}) + i k_{y} y}
\left( \begin{array}{c}
 \sqrt{V^{2}_1 - (\Delta^{2}_1 /4)} ~e^{\mp i \theta_1} \\
(V_1 + \Delta_1 /2) ~e^{\pm i \theta_1}
\end{array}
\right), \nonumber
\end{equation}
\end{widetext}
where $k_{1_x} = s_1 k_1 \cos \theta_1$,
$k_{y} = s_1 k_1 \sin \theta_1$, and
$s_1 = -\textrm{sgn} (V_1)$.
Here, $\theta_1$ can be regarded as the propagation angle of the plane wave in the $n$ region.
The setup in Fig.~4 of the main text satisfies the condition of $x_\textrm{src}  k_1 \gg 1$, hence the above approximation is well applicable to it;
%In our setup, the condition  $\rho |\vec{k}_1|~ \gg 1$ is valid,
%since the values of $\rho |\vec{k}_1|$ along the line of the junction
%($x = 0$) are larger than $|x_{{\rm src}} \vec{k}_1| = 22$ for $\Delta_1 / V_1 = -0.3$ and $V_1 = -0.02$;
we have checked that the error due to the approximation does not alter the main features discussed in the main text.
The approximation allows us to express
the refracted waves $\Psi^{{\rm K}({\rm K}')}_{{\rm rfr}}$ in terms of the transmission coefficient  $t_{np}(\theta_1)$ through the junction, which is mentioned in Eq.~(6) of the main text,
\begin{widetext}
\begin{equation}\label{RfrWave}
\Psi^{{\rm K}({\rm K}')}_{{\rm rfr}}(x,y)
\approx
\frac{-i}{\pi \sqrt{N_2}} \int^{\pi/2}_{-\pi/2} d \theta_1~
t_{np}(\theta_1)~e^{-i k_{1_x} x_{{\rm src}} + i k_{2_x} x + i k_{y} y}
\left( \begin{array}{c}
 \sqrt{V^{2}_2 - (\Delta^{2}_2 /4)} ~e^{\mp i \theta_2} \\
(V_2 + \Delta_2 /2) ~e^{\pm i \theta_2}
\end{array}
\right). \nonumber
\end{equation}
\end{widetext}
$k_{2_x},  N_2,$ and  $s_2$, are defined in the same way as $k_{1_x},  N_1,$ and $s_1$, except
for the subscript $2$.
%The upper (lower) sign corresponds to the K (K$'$) valley,
%$H^{(1)}_{1}$ is the Hankel function of the first kind,
%$(\rho~{\rm cos}~\phi, \rho~{\rm sin}~\phi) = (x + 100 l_0, y)$,
%$\hbar v k_{j_x} = s_j \sqrt{\gamma \sqrt{V^{2}_{j} - \Delta^{2}_{j}/4}}~\textrm{cos}~\theta_j$,
%$\hbar v k_{y} = s_1 \sqrt{\gamma \sqrt{V^{2}_{1} - \Delta^{2}_{1}/4}}~\textrm{sin}~\theta_1$,
%$N_j = 2 V^{2}_j + V_j \Delta_j$, $s_j = -\textrm{sgn} (V_j)$,
%$j \in \{ 1, 2 \}$ refers to the region with $V_j$,
%of $x < 0$ for $j = 1$ and $x > 0$ for $j = 2$,
The propagation angle $\theta_2$ of the refracted wave is governed by the conservation
of $p_y$, $s_1 \sqrt{\gamma \sqrt{V^{2}_{1} - \Delta^{2}_{1}/4}}~ \textrm{sin} \theta_1
= s_2 \sqrt{\gamma \sqrt{V^{2}_{2} - \Delta^{2}_{2}/4}}~ \textrm{sin} \theta_2$.
We calculate $t_{np}$ by using the boundary matching method of plane waves at $x = 0$, and then obtain the wave intensity from $I_{{\rm K}({\rm K}')}(x,y) = |\Psi^{{\rm K}({\rm K}')}_{{\rm rfr}}(x,y)|^2$. As we are considering the regime where the intervalley scattering is absent, the total intensity $I$ is the sum of $I_{\rm K}$ and $I_{\rm K'}$, i.e., there is no interference between the waves of K and K$'$ valleys.

Hereafter, we give additional information about the Veselago lens. We first mention the Klein effect in the smooth $np$ junction where $V(x)$ and $\Delta(x)$ vary smoothly around $x=0$.
In the main text, for the sharp junction in which the values of $V(x)$ and $\Delta(x)$ jump at $x=0$, we find the feature that $I(x,0)=0$ along the line of $y=0$ in the $p$ region of $x > 0$, when the condition of $\alpha_2 /\alpha_1 =1$ is achieved.
This feature robustly appears in the smooth $np$ junction.
 %where $V(x)$ and $\Delta(x)$ vary smoothly around $x=0$.
The robustness comes from the facts that the symmetry in Eq.~(5) of the main text is still valid, and that the value of $\Delta(x) / V(x)$ is almost constant~\cite{McCann_smooth} when $V(x)$ varies sufficiently smoothly.

Next we mention about electron propagation in the Veselago lens setup shown in Fig.~4 of the main text. In the $p$ region, the interference between three different refracted waves can appear at $(x \ge x_\textrm{cusp}, y=0)$; in contrast, there appears only a single refracted wave at $0 < x < x_\textrm{cusp}$. The three different waves propagate from the source to $(x,y=0)$ as follows.
One of them propagates from the source to the junction interface with incident angle $\theta_1 = \theta_x$ to the interface, and then it arrives at $(x,y=0)$ after its refraction at the interface. Here, the relation between $\theta_x$ and $x$ is found to be $x = x_{\rm cusp} \sqrt{(1-|V_1/V_2| {\rm sin}^2 \theta_x)/{\rm cos}^2 \theta_x}$, where $x_{\rm cusp} = |x_{\rm src}| \sqrt{|V_2/V_1|}$. Another wave follows the path with $\theta_1 = - \theta_x$, which is the mirror-reflection (about $y=0$ axis) path of the wave with $\theta_1 = \theta_x$. The other wave follows the path with normal incidence of $\theta_1 = 0$; the transmission amplitude of this wave through the interface of $x=0$ is in fact zero, when the condition of $\alpha_1 / \alpha_2 = 1$ for the Klein effect is achieved.
The interference between the three waves shows the clear signature of the Klein effect and the Berry phase $\pi$, as shown in the main text.

%Note that three refracted waves meet at ($x, 0$) when $x > x_{\rm cusp}$.
%In the classical trajectories of electrons shown in Fig.~4(a) in the main text,
%the refracted waves of incidence angle $\theta_1 = \theta_x$ (or $-\theta_x$) meets
%the line of $y=0$ at $x = x_{\rm cusp} \sqrt{(1-|V_1/V_2| {\rm sin}^2 \theta_x)/{\rm cos}^2 \theta_x}$,
%where $x_{\rm cusp} = |x_{\rm src}| \sqrt{|V_2/V_1|}$.
%Since the values of $x$ is larger than $x_{\rm cusp}$ for finite $\theta_x$,
%the three refraced waves of incidence angles $\theta_1 = 0, \theta_x,$ and $-\theta_x$
%interfere at ($x, 0$) when $x > x_{\rm cusp}$.

%In the region far from the source, $\rho |\vec{k}_1|~ \gg 1$,
%the plane-wave expansion \cite{Cincotti} of  is approximately given by
%We are interested in the $p$ region where the distance from the source is larger than
%$|x_{{\rm src}} \vec{k}_1| = 22$
%In our setup, we apply this approximation to the calcuation of the refractive waves in the $p$ region,
%We have checked that the
%Although the approximation gives the the approximation is correct within
%a good description of $\Psi^{{\rm K}({\rm K}')}_{{\rm src}}$ in the region where
%$\rho |\vec{k}_1|~ \gg 1$;
%In our setup, an interference $\rho |\vec{k}_1| \gtrsim |x_{{\rm src}} \vec{k}_1| = 22 ~ \gg 1$, and hence the
%approximation is valid.

%The method gives good approximation in our setup because the point source
%is far from the boundary of the junction and hence the validity of the approximation
%($100 l_0 \sqrt{k^{2}_{1_x} + k^{2}_{y}} \approx 22 \gg 1$ in our calculation) is satisfied.

\section{II. ~ Klein effect in armchair nanoribbon}

In the main text, we find that the Klein effect and the Berry phase $\pi$ can be detected in the Veselago lens.
In this section, we discuss another way of experimentally detecting the Klein effect
in a bilayer graphene armchair nanoribbon.
%We study electron transmission $t$ through the nanoribbon with an $npn$ junction.

The nanoribbon has armchair edges along $\hat{x}$ axis and shows metallic behavior in the absence of external potential and band gap. Its stacking configuration is shown in Supplementary Figure 1(a).
% and it has the width of $6 a$, where $a$ is the lattice constant of graphene.
We consider an $npn$ junction in the nanoribbon, formed by the potential energy $V(x)$ and band gap $\Delta(x)$,
\begin{equation} \label{Ribbon_potential}
(V(x),~\Delta (x)) = \left\lbrace
      \begin{array}{ll}
        (V_1 ,~\Delta_1), \,\,\, & \hbox{$x<0$ and $x > d$},\\
        (V_2,~\Delta_2), \,\,\, & \hbox{$0<x<d$}.
      \end{array}
    \right. \nonumber
\end{equation}
To see the signature of the Klein effect that the transmission probability of an electron through the junction is zero at normal incidence of $\theta_1 = 0$, one needs to study the energy regime having only one transverse mode, which corresponds to the state with $k_y=0$ in the bulk limit~\cite{Gonzalez}.
%The width of the nanoribbon of $6 a$, one transverse mode are in the energy range $\in [-\gamma, \gamma]$.
Focusing on this energy regime,
we numerically calculate the transmission probability $T = |t|^2$ through the $npn$ junction, by combining the tight-binding method and Green's function~\cite{Datta, Sim2}; as mentioned in the main text, $t$ is the transmission coefficient, and the energy of the incident electron is set by $E=0$.

\begin{figure}
\includegraphics[width=0.45\textwidth]{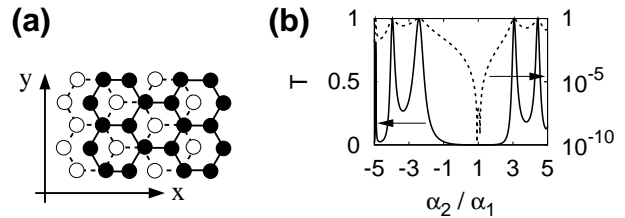}
\caption{Supplemantary figure 1.
(a) Stacking structure of a bilayer graphene armchair ribbon. Unfilled (filled) circles connected by dashed (solid) lines represent the lower (upper) layer of the bilayer ribbon. Electrons propagate along $\hat{x}$ direction.
(b) Transmission probability $T$ (with log scale plot on the right side) through an $npn$ junction
as a function of $\alpha_2 / \alpha_1$. The Klein effect of $T=0$ is shown at $\alpha_2 / \alpha_1 \approx 1$.
}
\label{fig:ribbon}
\end{figure}

In Supplementary figure 1(b), we plot $T$ as a function of $\alpha_2 / \alpha_1$, where $\alpha_{1(2)}= \Delta_{1(2)}/ V_{1(2)}$.
We choose $\alpha_1 = -0.3$, $V_1 = -0.05 \gamma$, $V_2 = 0.26 \gamma$,
$d = 60 l_0$, and the width of the ribbon $W = 6 a$, where $a$ is the lattice constant of graphene.
%; the same is shown for other values of $W$.
Here, the subindex $1$ ($2$) refers to the region of $x < 0$ and $x > d$ ($0 < x < d$).
We choose $V$ and $\Delta$ spatially varying over the length scale of $27 l_0$ around $x=0$ and $d$; in this case the intervalley mixing is prevented.

The transmission probability $T$ shows the perfect reflection of $T=0$ at $\alpha_2 / \alpha_1 \approx 1$. This shows the Klein effect, the transmission zero at normal incidence of $\theta_1=0$ in the bulk limit, hence, one can observe the Klein effect with tuning $\alpha_j$ by gate voltages.
This signature of the Klein effect (the perfect reflection) is distinguishable from the transmission valleys between resonance peaks. It is because the perfect reflection is independent of parameters such as $d/\lambda$, while the resonance peaks do depend on those parameters; $\lambda =  2 \pi \hbar v / \sqrt{\gamma \sqrt{V^{2}_{2} - \Delta^{2}_{2}/4}}$ is an electron wavelength in the $p$ region.

%Although the small change of the length scale causes the shift of resonance centers (peak positions of $T$) and of transmission valleys between the centers,

The Klein effect in the armchair nanoribbon is robust against the details of the shapes
of $V(x)$ and $\Delta (x)$ around the interfaces of the $npn$ junction, provided that the valley mixing is negligible.
The robustness implies that the Klein effect
will survive in the presence of screened-Coulomb interaction, because the interaction
may induce small change of the spatial shape of $V(x)$ and $\Delta (x)$ at the interfaces \cite{Zhang}.

The above finding appears in a metallic armchair ribbon with arbitrary width, provided that the energy
regime has one transverse mode corresponding to the states with $k_y = 0$ in the bulk limit.
Note that it is difficult to observe the effect in ribbons with non-metallic armchair edges or with zigzag edges.
This is because there is no state corresponding to the $k_y = 0$ state injecting normal to the junction, i.e., because the band dispersion of the ribbon in the low-energy regime is different
from that in bulk limit due to the edges.

The perfect reflection may disappear in the presence of valley-symmetry-breaking disorders, e.g.,
short range disorders and edge disorders.
Such disorders induce valley mixing, and hence the Berry phase argument based on the time reversal and
electron-hole symmetries defined in a single valley is not applicable.

%depends only on the pseudospin polarization vectors
%in the $n$ and $p$ regions, regardless of other details such as the varying shape of the potential
%and the gap.

\end{document}